\documentclass[prb,superscriptaddress,twocolumn]{revtex4-1}
\usepackage[utf8]{inputenc}
\usepackage{amsmath}
\usepackage{amsfonts}
\usepackage{amssymb}
\usepackage{graphicx}
\usepackage{lipsum}
\usepackage{hyperref}
\usepackage{xcolor}

\begin{document}

\title{Magnetic order, hysteresis and phase coexistence in magnetoelectric LiCoPO$_4$}
\date{\today}

\author{Ellen Fogh}
\affiliation{Department of Physics, Technical University of Denmark, DK-2800 Kongens Lyngby, Denmark}
\author{Rasmus Toft-Petersen}
\affiliation{Department of Physics, Technical University of Denmark, DK-2800 Kongens Lyngby, Denmark}
\author{Eric Ressouche}
\affiliation{INAC-SPSMS, CEA \& Universite Grenoble Alpes, 38000 Grenoble, France}
\author{Christof Niedermayer}
\affiliation{Laboratory for Neutron Scattering and Imaging, Paul Scherrer Institute, Villigen CH-5232, Switzerland}
\author{Sonja Lindahl Holm}
\affiliation{Laboratory for Neutron Scattering and Imaging, Paul Scherrer Institute, Villigen CH-5232, Switzerland}
\affiliation{Nano-Science Center, Niels Bohr Institute, University of Copenhagen, DK-2100 Copenhagen Ø, Denmark}
\author{Maciej Bartkowiak}
\affiliation{Helmholtz-Zentrum Berlin für Materialien und Energie, D-14109 Berlin, Germany}
\author{Oleksandr Prokhnenko}
\affiliation{Helmholtz-Zentrum Berlin für Materialien und Energie, D-14109 Berlin, Germany}
\author{Steffen Sloth}
\affiliation{Department of Physics, Technical University of Denmark, DK-2800 Kongens Lyngby, Denmark}
\author{Frederik Werner Isaksen}
\affiliation{Department of Physics, Technical University of Denmark, DK-2800 Kongens Lyngby, Denmark}
\author{David Vaknin}
\affiliation{Ames Laboratory and Department of Physics and Astronomy, Iowa State University, Ames, Iowa 50011}
\author{Niels Bech Christensen}
\affiliation{Department of Physics, Technical University of Denmark, DK-2800 Kongens Lyngby, Denmark}

\begin{abstract}

The magnetic phase diagram of magnetoelectric LiCoPO$_4$ is established using neutron diffraction and magnetometry in fields up to $25.9\,\mathrm{T}$ applied along the crystallographic $b$-axis. For fields greater than $11.9\,\mathrm{T}$ the magnetic unit cell triples in size with propagation vector ${\bf Q} = (0,1/3,0)$. A magnetized elliptic cycloid is formed with spins in the $(b,c)$-plane and the major axis oriented along $b$. Such a structure allows for the magnetoelectric effect with an electric polarization along $c$ induced by magnetic fields applied along $b$. Intriguingly, additional ordering vectors ${\bf Q} \approx (0,1/4,0)$ and ${\bf Q} \approx (0,1/2,0)$ appear for increasing fields in the hysteresis region below the transition field. Traces of this behavior are also observed in the magnetization. A simple model based on a mean-field approach is proposed to explain these additional ordering vectors. In the field interval $20.5-21.0\mathrm{T}$, the propagation vector ${\bf Q} = (0,1/3,0)$ remains but the spins orient differently compared to the cycloid phase. Above $21.0\,\mathrm{T}$ and up until saturation a commensurate magnetic structure exists with a ferromagnetic component along $b$ and an antiferromagnetic component along $c$.

\end{abstract}

\maketitle

\section{Introduction}

The rich physics of magnetically frustrated systems -- although studied theoretically and experimentally for half a century\cite{ramirez1994,balents2010} -- continues to attract interest in condensed matter research. Frustration is either imposed by the geometry of the spin lattice or caused by competing interactions. In either case, the system can not minimize all interaction energies simultaneously. One possible outcome is a non-ordered and highly degenerate state where the system fluctuates between many different configurations down to very low temperatures. In this case, one encounters exotic materials such as spin ices\cite{harris1997} and quantum spin liquids\cite{anderson1973} displaying magnetic monopoles and spinon excitations. Alternatively, the system finds a stable minimum energy configuration which often brings about non-collinear and/or incommensurate magnetic structures\cite{kimura2007}. In turn, these structures are closely linked to multiferroicity, magnetostriction and magnetoelectricity \cite{kimura2007} -- just to mention a few profound curiosities of technological and fundamental interest. The symmetries of the nuclear and magnetic structures govern how the individual material properties are manifested. Finally, in combination with disorder, frustrated interactions may ultimately result in spin glasses where spins are frozen in at random\cite{moorjani1978}.

The lithium orthophosphates, Li$M$PO$_4$ ($M$ = Co, Ni, Mn, Fe), are a family of compounds with orthorhombic symmetry (space group \textit{Pnma}) which all exhibit commensurate antiferromagnetism as well as the magnetoelectric effect in their ground states\cite{santoro1966,wiegelmann}. In these materials, the coupling between ferroelectricity and antiferromagnetism is governed by the magnetic structure\cite{wiegelmann,mercier}, the details of which are also believed explain the effect in LiCoPO$_4$\cite{vaknin2002,kornev2000}. Previously it has been shown that the magnetoelectric effect in LiNiPO$_4$ is closely related to a field-induced spin canting faciliated by the Dzyaloshinskii-Moriya interaction\cite{jensen2009_2}. The magnetoelectric effect in LiCoPO$_4$ is by far the strongest in the lithium orthophosphate family\cite{wieglhofer} but the microscopic mechanism behind it is yet to be understood.

LiCoPO$_4$ has cell parameters $a = 10.20\,\mathrm{Å}$, $b = 5.92\,\mathrm{Å}$ and $c = 4.70\,\mathrm{Å}$\cite{newnham1965} and the four magnetic Co$^{2+}$ ions ($S=3/2$) of the crystallographic unit cell form an almost face-centered structure with the positions ${\bf r}_1 = (1/4+\varepsilon, 1/4, 1-\delta)$, ${\bf r}_2 = (3/4+\varepsilon, 1/4, 1/2+\delta)$, ${\bf r}_3 = (3/4-\varepsilon, 3/4, \delta)$ and ${\bf r}_4 = (1/4-\varepsilon, 3/4, 1/2-\delta)$ where $\varepsilon = 0.0286$ and $\delta = 0.0207$\cite{kubel1994}. The displacement, $\varepsilon$, of the ions gives rise to a toroidal moment as demonstrated both theoretically\cite{ederer2007,spaldin2008} and experimentally\cite{vanAken2007,zimmermann2014}. The easy axis for the Co$^{2+}$ ions is along $b$ as deduced from the magnetic susceptibility\cite{creer1970}. Here the moments are magnetized twice as readily along $b$ as compared to along $a$ and $c$ in the paramagnetic phase. Furthermore, although the susceptibilities along $a$ and $c$ are of similar magnitude, $a$ is the harder axis. A priori density functional theory calculations agree with these measurements\cite{yamauchi2010}. Hence, the single-ion anisotropy of the Co$^{2+}$ ions is largely axial and with the easy axis along $b$. Below $T_N = 21.6\,\mathrm{K}$ LiCoPO$_4$ orders antiferromagnetically with spins along the easy axis in a commensurate $(\uparrow \uparrow \downarrow \downarrow)$ arrangement\cite{szewczyk2011,santoro1966}. Here $\uparrow$ and $\downarrow$ denote spin up and down respectively for the ion sites in forthcoming order, i.e. in the above case spins 1 and 2 are up and spins 3 and 4 are down. In addition, a small spin rotation away from the $b$-axis as well as a weak ferromagnetic moment have been reported\cite{vaknin2002,rivera1994,kharchenko2003}.

Pulsed-field magnetic susceptibility measurements up to $29\,\mathrm{T}$ at liquid He temperatures show a number of phase transitions\cite{kharchenko2010}. At $\sim12\,\mathrm{T}$, the magnetization jumps to a plateau of $1/3$ of its saturation value, $M_S = 3.6\mu_B$/Co-ion. Next, at $\sim22\,\mathrm{T}$, it gradually increases to $2/3\,M_S$ and then finally increases linearly until saturation is achieved at $\mu_0 H_S = 28.3\,\mathrm{T}$. The magnetolectric tensor element $\alpha_{ab}$ was recently probed in a pulsed-field electric polarization experiment ($P_i = \alpha_{ij} H_j$ where $i,j = \lbrace a,b,c \rbrace$, $P_i$ is the induced electric polarization and $H_j$ the applied magnetic field). The measurements show that the phase in the interval $22-28\,\mathrm{T}$ supports the magnetoelectric effect but with considerably smaller magnetoelectric coefficient compared to the commensurate low-field phase\cite{khrustalyov2016}. The intermediate phase displaying the $1/3$ magnetization plateau does not, on the other hand, exhibit the magnetoelectric effect for this coefficient.

The magnetic exchange interactions of LiCoPO$_4$ are shown in Fig. \ref{fig:phasediagram} together with the magnetic unit cell of the commensurate low-field structure. The interactions in the lithium orthophosphates are generally frustrated leading to a multitude of phases as a function of temperature and applied magnetic field\cite{toftpetersen2011,toftpetersen2012,toftpetersen2017}. The nearest neighbor interaction, $J_{bc}$, is antiferromagnetic but so are $J_b$ and $J_{ab}$ (terminology adobted from Ref. \onlinecite{jensen2009}). The interactions $J_{c}$ and $J_{ac}$ are weak and may differ in sign depending on the magnetic ion in question\cite{toftpetersen2015}. The exchange interactions are mediated via super-exchange paths such as $M$-O-$M$ or $M$-O-P-O-$M$\cite{szewczyk2011,vaknin2002,kornev2000}. In Ref. \onlinecite{kharchenko2010}, the values of $J_{bc}$, $J_b$ and $J_c$ were estimated for LiCoPO$_4$ from the transition field values using a model for the magnetic structures based on magnetization measurements exclusively. Collinear structures with moments along $b$ and propagation vector along $c$ were assumed in all phases. In a different study, the spin wave spectrum was measured and although the fitted exchange parameters are subject to large uncertainties they offer a reasonable estimate for the interactions\cite{tian2008} \footnote{We have examined crystals from the same batch as that used in Ref. [30] and find a significantly lower transition temperature, $T_N = 17.3(1)\,\mathrm{K}$. Furthermore, a Rietveld refinement of our neutron diffraction data yields satisfactory results exclusively when introducing Ni as well as Co on the magnetic site. Hence, our results suggest that the crystals may not be pure LiCoPO$_4$, but possibly Ni-doped from a crucible-growth.}.

In the present work we investigate the phase diagram of LiCoPO$_4$ up to $25.9\,\mathrm{T}$ for magnetic fields applied along $b$. We present magnetization and neutron diffraction results for the field-induced transition at $11.9\,\mathrm{T}$. These provide direct evidence that the ordering vector of the phase with $1/3$ magnetization is ${\bf Q} = (0,1/3,0)$ and the spin arrangement is a superposition of a cyloid structure in the $(b,c)$-plane and a ferromagnetic component. Furthermore, hysteresis is observed as well as pronounced differences in the way the transition occurs depending on field ramp direction. For increasing field several magnetic Bragg peaks signifying different incommensurate spin structures coexist in the region below the transition, $11.4-11.9\,\mathrm{T}$. For decreasing field the transition appears abruptly but for a broadening of the commensurate peak at the transition. We also present neutron diffraction results for the phases at $20.5-21.0\,\mathrm{T}$ and above $21.0\,\mathrm{T}$. The former has propagation vector ${\bf Q} = (0,1/3,0)$ too but a different spin orientation compared to the cycloid phase. The latter is commensurate, most likely with a ferromagnetic component along $b$ as well as an antiferromagnetic component along $c$.

\section{Experimental details}

Magnetization measurements were carried out using the vibrating sample method with a standard CRYOGENIC cryogen free measurement system. Magnetic fields of $0 \leq \mu_0 H \leq 16\,\mathrm{T}$ were applied along the $b$-axis for temperatures in the interval $2\,\mathrm{K} \leq T \leq 300\,\mathrm{K}$. 

Neutron diffraction experiments were performed at the triple-axis spectrometer RITA-II at the Paul Scherrer Institute with a PG(002) vertically focussing monochromator and 80' collimation between monochromator and sample. The instrument was operated with incoming and outgoing wavelength $\lambda = 4.04\,\mathrm{Å}$ and a cooled Be filter before the analyzer. Vertical magnetic fields up to $15\,\mathrm{T}$ were applied along $b$ and momentum transfers were confined to the $(H,0,L)$ plane.

Studies with magnetic fields up to $12\,\mathrm{T}$ along the $b$-axis took place at the diffractometer D23 at the Institute Laue-Langevin utilizing neutrons of wavelength $\lambda = 1.279\,\mathrm{Å}$ and with no collimation. A lifting detector and cryomagnet with asymmetric opening angles allowed for measurements of momentum transfers with an out-of-plane component. This proved pivotal for identifying the propagation vector along $b$. 86 commensurate peaks were collected at $(30\,\mathrm{K},0\,\mathrm{T})$, $(2\,\mathrm{K},0\,\mathrm{T})$ and $(2\,\mathrm{K},12\,\mathrm{T})$ and 91 incommensurate peaks were collected at $(2\,\mathrm{K},12\,\mathrm{T})$. Circular diaphragms of $15\,\mathrm{mm}$ and $6\,\mathrm{mm}$ were used for peak collection and scans along $(3,K,1)$ respectively.

Further measurements were performed at the high magnetic field facility for neutron scattering, which consists of the Extreme Environment Diffractometer (EXED) and a High Field Magnet (HFM), at the Helmholtz-Zentrum Berlin\cite{smeibidl2016,prokhnenko2015,prokhnenko2016}. This truly unique horizontal hybrid solenoid magnet allowed to directly probe all magnetic phases up to $25.9\,\mathrm{T}$ DC field. The magnet has $30^{\circ}$ conical openings, which combined with magnet rotation with respect to the incident neutron beam and the time-of-flight neutron technique implemented on EXED, allows access to a substantial region of reciprocal space. In our case the crystal was oriented with $(0,1,0)$ and $(1,0,1)$ in the horizontal scattering plane and magnetic fields were applied along the $b$-axis with temperatures in the range $1.1-30\,\mathrm{K}$. Two different magnet and EXED chopper settings were employed for measuring Bragg peaks occurring in the forward scattering and backscattering detectors respectively: (i) magnet rotation $-11.83^{\circ}$ with respect to the incoming beam, wavelength band $0.7-1.7\,\mathrm{Å}$ (wavelength resolution $\sim 4-2$\%) and (ii) magnet rotation $-10.5^{\circ}$, wavelength band $4.8-12.0\,\mathrm{Å}$ (wavelength
resolution $\sim 0.6-0.2$\%)

The same high quality LiCoPO$_4$ single crystal measuring $\sim2\times2\times5\,\mathrm{mm^3}$ ($21.4\,\mathrm{mg}$) was used for both magnetization measurements and neutron diffraction experiments. In all cases, the crystal was aligned such that $H||b$ within about $1^{\circ}$ except at HFM/EXED where the alignment was within $3^{\circ}$.

\section{Results}

\subsection{Phase diagram}

\begin{figure}[b!]
	\centering
	\includegraphics[width = \columnwidth]{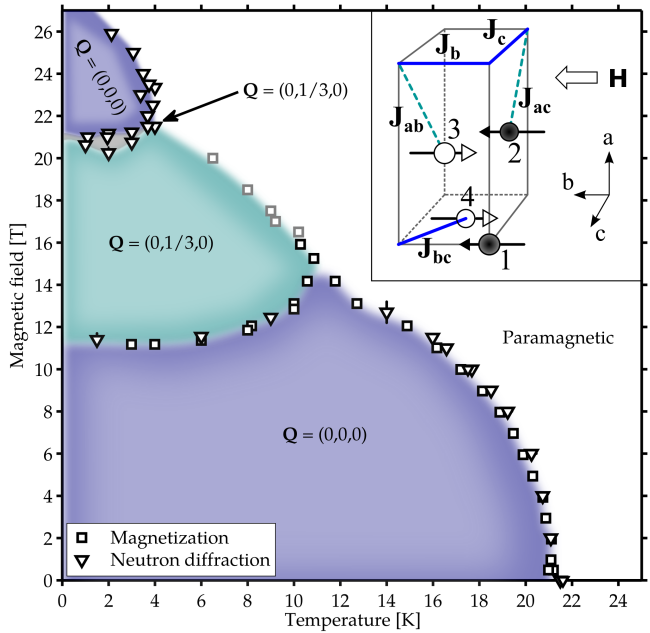}
	\caption{Magnetic phase diagram of LiCoPO$_4$ for fields up to $25.9\,\mathrm{T}$ applied along $b$ as measured by magnetization (square symbols) and neutron diffraction (triangular symbols). The grey symbols are from Ref. \onlinecite{wiegelmann}. The transition fields and temperatures are determined for increasing fields and upon cooling below $16\,\mathrm{T}$ and upon heating above. The propagation vectors identified from neutron diffraction are shown in the respective phases. The zero-field magnetic unit cell with exchange interactions indicated is shown as an insert. Note that only the major spin component along $b$ is shown, see text.}
	\label{fig:phasediagram}
\end{figure}

\begin{figure}[t!]
	\centering
	\includegraphics[width = \columnwidth]{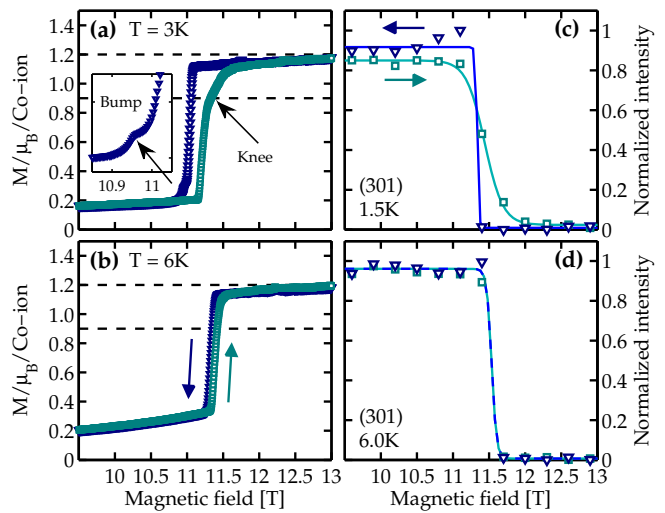}
	\caption{Magnetization and neutron diffraction data from RITA-II as a function of field for both increasing (green squares) and decreasing (blue triangles) field. (a) and (b) show the magnetization at $3\,\mathrm{K}$ and $6\,\mathrm{K}$ respectively. The material exhibits hysteresis and the insert and arrows emphasize the special features at $3\,\mathrm{K}$ discussed in section III.C. These features are absent at $6\,\mathrm{K}$. The dashed lines indicate $1/3$ and $1/4$ of the saturation magnetization. (c) and (d) show the integrated intensity of the magnetic Bragg peak $(3,0,1)$ at $1.5\,\mathrm{K}$ and $6\,\mathrm{K}$ respectively. The solid lines are guides to the eye. Blue and green arrows indicate the field ramp directions.}
	\label{fig:mag_rita}
\end{figure}

Using magnetization measurements and by tracking the temperature and field dependencies of selected magnetic Bragg peaks, the magnetic phase boundaries of LiCoPO$_4$ were determined for fields up to $25.9\,\mathrm{T}$ applied along $b$, see Fig. \ref{fig:phasediagram} and Fig. \ref{fig:mag_rita}. The boundary as a function of temperature for fields in the interval $16-20\,\mathrm{T}$ is reproduced from Ref. \onlinecite{wiegelmann} and fits well with our results. The transition temperature in zero field is found to be $T_N = 21.6(1)\,\mathrm{K}$ in good agreement with literature and the refined zero-field structure is consistent with the $(\uparrow \uparrow \downarrow \downarrow)$ structure with spins along $b$ as previously reported by others\cite{szewczyk2011,santoro1966,vaknin2002}. Field-induced phase transitions were observed at $11.9$, $20.5$, and $21.0\,\mathrm{T}$ at liquid He temperatures, also largely in agreement with previous pulsed-field magnetization results\cite{kharchenko2010}. The Curie-Weiss temperature, $\theta_{CW} = 121(1)\,\mathrm{K}$, was determined from the inverse magnetic susceptibility (not shown) at $0.5\,\mathrm{T}$ applied along $b$. Thus the frustration parameter\cite{ramirez1994}, $f = \frac{\theta_{CW}}{T_N} \approx 5$, indicates the presence of moderate frustation in the system. Furthermore, the shape of the phase boundary as a function of temperature is somewhat unusual with the transition temperature being considerably supressed at $12\,\mathrm{T}$ compared to zero field, $T_N(H = 1/3 \, H_S) \approx 1/2 \, T_N(H = 0)$ and to an even greater extent at $21\,\mathrm{T}$, $T_N(H = 3/4 \, H_S) \approx 1/5 \, T_N(H = 0)$. In contrast, for the sister compound, LiNiPO$_4$, an incommensurate phase exists at higher temperatures for fields up to $17.3\,\mathrm{T}$\cite{jensen2009_2}. In LiCoPO$_4$, however, no phase transition is observed above $10\,\mathrm{K}$ at $16\,\mathrm{T}$, neither in the magnetization nor heat capacity (not shown here). This also explains why at $14\,\mathrm{K}$ no magnetoelectric effect was observed as a function of field above the commensurate low-field phase in Ref. \onlinecite{khrustalyov2016}. Although peculiar compared to the sister compounds, the shape of the phase boundary is similar to that found in other Co$^{2+}$ Ising systems such as BaCo$_2$V$_2$O$_8$\cite{kimura2008}. This is true even if crystal structure and single-ion anisotropies differ greatly from those of LiCoPO$_4$.

\subsection{Magnetic structure with 1/3 magnetization}

\begin{figure}[b!]
	\includegraphics[width = \columnwidth]{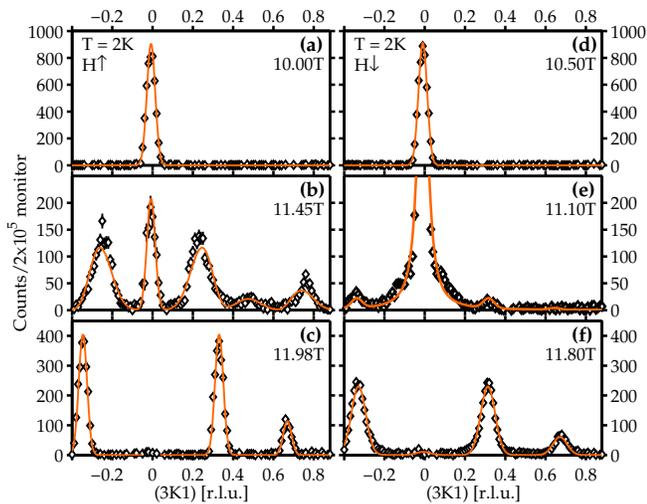}
	\caption{Neutron scattering intensity as a function of $(3,K,1)$ for selected fields at $2\,\mathrm{K}$ for (a)-(c) increasing and (d)-(f) decreasing field. The selected fields are: in the low-field commensurate phase (top panels), in the transition region (middle panels) and in the ${\bf Q} = (0,1/3,0)$ phase (bottom panels). The actual field values are given in the plots. The solid lines are fits to the respective data set as described in the text.}
	\label{fig:lineshapes}
\end{figure}

\begin{figure}[t!]
	\includegraphics[width = \columnwidth]{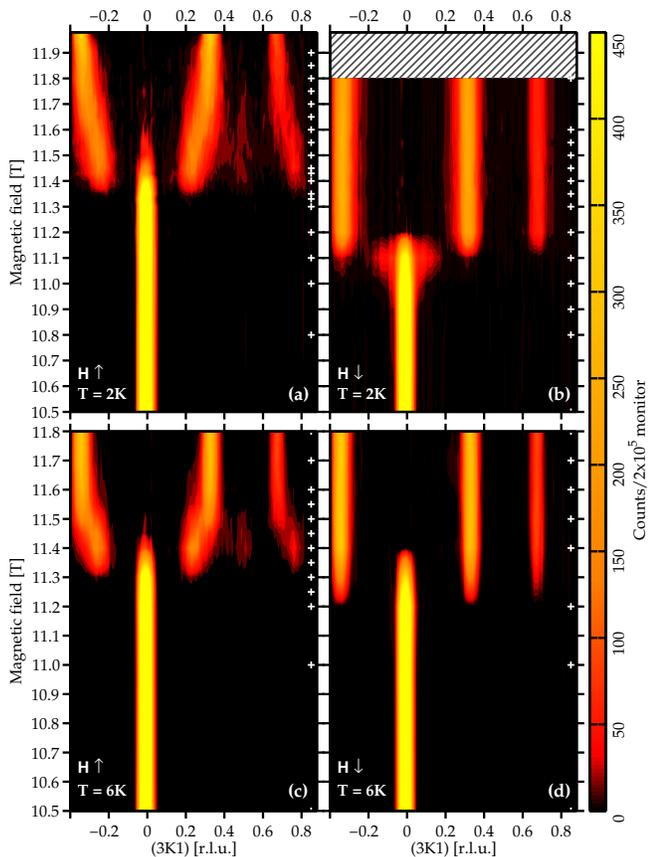}
	\caption{Colorplots of the intensity of $(3,K,1)$ as a function of magnetic field applied along $b$ at (a)-(b) $2\,\mathrm{K}$ and (c)-(d) $6\,\mathrm{K}$ for both increasing and decreasing field as measured at D23. The white crosses to the right in each colorplot denote the field values for which scans have been performed. Note the relatively few points in (d) and the difference in maximum field between the top and bottom panels. No data was collected in the hatched area.}
	\label{fig:colorplots}
\end{figure}

Neutron diffraction was employed to determine the magnetic structure in the field interval $11.9-20.5\,\mathrm{T}$ with $1/3$ magnetization. Having observed the disappearance of the Bragg peaks characteristic of the commensurate low-field phase, see Fig. \ref{fig:mag_rita}(c) and (d), we searched extensively for Bragg peaks in the $(H,0,L)$ scattering plane but with no success. Hence the ordering vector is neither along $a$ nor $c$ nor a number of superpositions of those two directions. Furthermore, in the sister compound LiNiPO$_4$ the ordering vector is $(0,K,0)$ with $K$ attaining both rational and irrational values depending on field and temperature\cite{toftpetersen2011}. It is therefore tempting to infer that the propagation vector is along $b$ for LiCoPO$_4$ too. However, with the field applied vertically also along this direction one needs a magnet with a sufficiently large opening angle and a detector with the ability to measure momentum transfers with finite out-of-plane components. Fortunately, D23 at the Institute Laue-Langevin offers such a setup. Scans of $(3,K,1)$ were performed at various field strengths and for increasing and decreasing field. Fig. \ref{fig:lineshapes} shows intensity profiles as a function $K$ along the $(3,K,1)$ direction at selected field values. Figure \ref{fig:colorplots} shows colorplots produced from a series of such scans performed at $2\,\mathrm{K}$ and $6\,\mathrm{K}$. The ordering vector of the structure was determined to be ${\bf Q} = (0,0.33(1),0) \approx (0,1/3,0)$ based on Gaussian fits to the observed resolution limited incommensurate peaks along $(3,K,1)$ at $2\,\mathrm{K}$ and $11.98\,\mathrm{T}$, \textit{cf.} Fig. \ref{fig:lineshapes}(c). Consequently, the magnetic unit cell triples along the crystallographic $b$-direction.

\begin{table}[t!]
	\centering
	\caption{Atomic positions for LiCoPO$_4$ obtained from Fullprof refinement ($R_F = 5.23$\%) of 86 commensurate peaks collected at D23 at $(30\,\mathrm{K}, 0\,\mathrm{T})$ and using the \textit{Pnma} space group. The Debye-Waller factor was refined globally to $B_{\mathrm{iso}} = 0.08$ and a Becker-Coppens type extinction correction has been applied. Fourier components for the cycloid formed by the magnetic Co$^{2+}$ ions are given in the two rightmost columns. These were refined ($R_F = 11.1$\%) from 91 incommensurate peaks collected at $(2\,\mathrm{K},12\,\mathrm{T})$. $R_m$ and $I_m$ denote the real and imaginary Fourier coefficient respectively. These correspond to the moment sizes in $\mu_B$ along the major and minor axes of the enveloping ellipsoid.}
	\label{tab:fullprof}
    \begin{ruledtabular}
    \begin{tabular}{c c c c c c c}
    	Atom	& Site	& $x$		& $y$		& $z$			& $R_m$		& $I_m$	\\
    	\hline
	    	Li 		& 4a	& 0			& 0			& 0			& --			& --\\
		Co 		& 4c	& 0.2771(9) 	& 0.25		& 0.980(3)	& 4.13(5)	& 1.3(2)\\
		P		& 4c	& 0.0951(6) 	& 0.25		& 0.414(1)	& --			& --\\
		O1		& 4c	& 0.0975(4) 	& 0.25 		& 0.744(1)	& --			& --\\
		O2		& 4c	& 0.4542(4) 	& 0.25 		& 0.208(1)	& --			& --\\
		O3		& 8d	& 0.1663(2) 	& --  		& 0.2814(5)	& --			& --
    \end{tabular}
    \end{ruledtabular}
\end{table}

\begin{figure*}
	\centering
	\includegraphics[width = \textwidth]{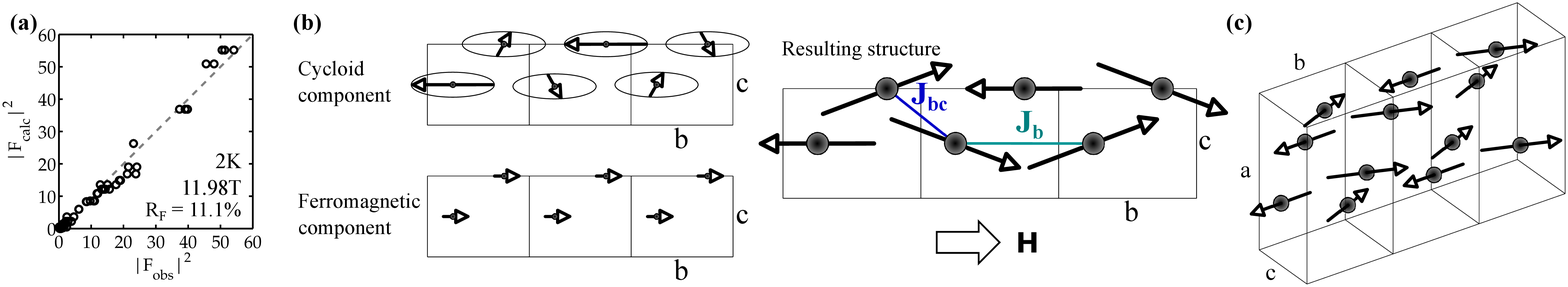}
	\caption{Refined magnetic structure. (a) calculated vs. observed scattering intensities for the collected incommensurate peaks as obtained in Fullprof for the refined magnetic structure. The dashed line shows $|F_{\mathrm{calc}}|^2 = |F_{\mathrm{obs}}|^2$. (b) and (c) magnetic structure for $11.9-20.5\,\mathrm{T}$ applied along $b$ shown in the $(b,c)$-plane and in 3D respectively. The spins order in a superposition of an elliptic cycloid and a ferromagnetic component along $b$. This results in 2/3 of the spins being almost parallel and 1/3 being antiparallel to the field direction. In (b) the nearest and next-nearest neighbor interactions, $J_{bc}$ and $J_b$, are shown.}
	\label{fig:structure}
\end{figure*}

From the 91 incommensurate peaks collected at $(2\,\mathrm{K}, 12\,\mathrm{T})$, an elliptic cycloid structure was refined using FullProf\cite{rodriguez-carvajal1993}. Here all spins with the same $y$ coordinate align and form a layer in the $(a,c)$-plane. Spins in subsequent layers rotate $\sim120^{\circ}$ in the $(b,c)$-plane upon advancing along the $b$-axis. The ratio between the major and minor axes of the enveloping ellipse is $3.2(5)$ with the major axis along $b$. The calculated versus observed intensities are shown in Fig. \ref{fig:structure}(a). Refinement results for the crystal structure and Fourier components of the magnetic structure are given in Table \ref{tab:fullprof}.

The $1/3$ magnetization implies an additional ferromagnetic component to be combined with the incommensurate structure. For the cycloid part of the structure there is as always an indeterminable phase shift which in this case has been set to $\pi/3$. This choice maximizes all spin lengths and allows $1/3$ of the spins to be along the easy $b$-axis. The energy cost associated with the single-ion anisotropy is independent of the phase shift angle. Assuming $M_S = 3.6\mu_B$\cite{kharchenko2010} and choosing the phase shift to $\pi/3$ the cycloid and ferromagnetic components result in the structure illustrated in Fig. \ref{fig:structure}(b) and (c).

\subsection{Hysteresis and phase coexistence}

Hysteresis is observed both in the magnetization measurements and in the neutron diffraction data at the transition from the low-field collinear phase to the magnetized cycloid phase. At a first order transition, one expects hysteresis to be present, but the case of LiCoPO$_4$ the transition is accompanied by additional field ramp direction dependent characteristics. How this is manifested is described in the following paragraphs.

In the field scans of the magnetization shown in Fig. \ref{fig:mag_rita}(a)-(b) hysteresis is present at $3\,\mathrm{K}$ but significantly reduced at $6\,\mathrm{K}$. Furthermore, at $3\,\mathrm{K}$ the shape of the magnetization curve depends on the field ramp direction as follows: for increasing field, the transition is first abrupt with the magnetization jumping to $\sim1/4 \, M_S$. Hereafter it increases approximately linearly until the $1/3$ magnetization plateau is reached. Conversely, for decreasing field, the transition is abrupt but the magnetization exhibits a minor bump before the system finally enters the low-field phase. At $6\,\mathrm{K}$ only minimal hysteresis is observed and the magnetization curves for increasing and decreasing field are similar to each other with just a single step from the low-field phase to $1/3 \, M_S$.

Correspondingly, field scans of the strong $(3,0,1)$ magnetic peak measured by neutron diffraction are shown in Fig. \ref{fig:mag_rita}(c)-(d). At $1.5\,\mathrm{K}$ a transition initiates at $\sim 11.4\,\mathrm{T}$ as a function of increasing field strength, in good agreement with earlier findings\cite{kharchenko2010}. For decreasing field, the transition appears at a somewhat lower field, $\sim 11.3\,\mathrm{T}$. Again the curve follows different trends depending on the field ramp direction: for increasing field the transition appears smooth whereas for decreasing field it is abrupt. At $6\,\mathrm{K}$ both hysteresis and other ramp direction dependent behavior are absent. The slight differences in the observed transition fields when comparing neutron diffraction data and magnetization measurements may be explained by differences in temperature.

Likewise, hysteresis of about $0.3\,\mathrm{T}$ is evident when comparing Fig. \ref{fig:colorplots}(a) and (b). For increasing field, the transition commences around $11.4\,\mathrm{T}$ where the intensity of the commensurate $(3,0,1)$ Bragg peak begins to decrease and peaks appear at $\sim(3,\pm 0.2,1)$. Upon further increasing the field they appear to gradually move to $(3,\pm0.33,1)$ where they lock in at $\sim 11.9\,\mathrm{T}$. In addition, a less intense peak is observed at $\sim(3,0.5,1)$ in the transition region, $11.4\,\mathrm{T}-11.9\,\mathrm{T}$, where some intensity is also still present at the commensurate position. In this region the incommensurate peaks are broadened and their shapes are asymmetric as can be seen by comparing Fig. \ref{fig:lineshapes}(b) and (c). Above $11.9\,\mathrm{T}$ the peaks become resolution limited and symmetric.

One possible explanation for the observed behavior is incommensurate order with a field-dependent unit cell size. However, such long range order would result in resolution limited symmetric peaks and can therefore be ruled out. The peak broadening indicates finite domain sizes and the line shape asymmetry may find its origin in overlapping peaks, possibly signifying several structures with different propagation vectors. The seemingly changing peak position may then be attributed to the change in volume ratio between the different structures involved.

The fit to the $11.45\,\mathrm{T}$ scan shown in Fig. \ref{fig:lineshapes}(b) is based on a model with ordering vectors ${\bf Q} = (0,1/3,0)$, $(0,1/4,0)$ and $(0,1/2,0)$. While the $(0,1/3,0)$ propagation vector is kept fixed at the value found at $11.98\,\mathrm{T}$, the other two are fitted globally to all data sets in the transition region. The peak intensities are allowed to vary between data sets but the intensities of the two peaks in a pair, $(3,\pm K,1)$, are kept equal. The globally fitted propagation vectors are  $(0,0.26(1),0)\approx(0,1/4,0)$ and $(0,0.48(3),0) \approx (0,1/2,0)$. Several other models were considered, including one involving an additional ordering vector ${\bf Q} = (0,1/5,0)$ and another where the incommensurate peaks were fitted to a single but field-dependent position. Neither of these were successful.

The observation of several propagation vectors in the transition region suggests a substantial degree of frustration and the existence of a number of spin configurations with only small energy differences. Steps in the magnetization accompanied by magnetic structures of rational periods, the so-called devil's staircase, are characteristica of the axial Ising antiferromagnet\cite{bak1986}. Even though LiCoPO$_4$ is not an entirely adequate model material for an Ising system, its spin configurations still seem to occur with rational periods. Hence such behavior may not be limited to the strict Ising case.

For decreasing field the transition proceeds entirely differently. Upon decreasing the field, the incommensurate $(3,1/3,1)$ peak abruptly gives way to the commensurate $(3,0,1)$ peak at $11.1\,\mathrm{T}$, consistent with RITA-II and magnetization data, compare Fig. \ref{fig:colorplots}(b) and Fig. \ref{fig:mag_rita}(c). Note that the incommensurate peaks are wider for decreasing field than the resolution limited peak measured at $11.98\,\mathrm{T}$ for increasing field. This is likely due to the fact that for decreasing field, the field was only ramped to $11.8\,\mathrm{T}$ before starting the measurements. The peak widths at $11.8\,\mathrm{T}$ in Fig. \ref{fig:colorplots}(a) and (b) are equal within the error of the fit. In the picture with separate domains with ordering vectors ${\bf Q} =(0,1/3,0)$ and $(0,1/4,0)$, the system is trapped in the $11.8\,\mathrm{T}$ state. This is below $11.9\,\mathrm{T}$ where the peaks become resolution limited and the structure is described purely by ${\bf Q} = (0,1/3,0)$.

At $11.1\,\mathrm{T}$, the commensurate $(3,0,1)$ peak is broadened and has a Lorentzian line shape, indicating disorder, see Fig. \ref{fig:lineshapes}(e). Fitting a Lorentzian convoluted with a Gaussian describing the resolution, one can obtain the correlation length as $\xi = \frac{b}{2 \pi \kappa}$, where $b$ is the lattice parameter and $\kappa$ is the Lorentzian width. The resolution is found by fitting the commensurate peak at $10.5\,\mathrm{T}$ (well below the transition) to a Gaussian, see Fig. \ref{fig:lineshapes}(d). The correlation length is then found to be $\sim 120$ times smaller just at the transition ($11.1\,\mathrm{T}$) compared to below ($10.5\,\mathrm{T}$). The observed peak broadening correlates with the bump seen in the magnetization, see Fig. \ref{fig:mag_rita}(a).

\subsection{Magnetic structures at high fields}

\begin{figure}[b!]
	\includegraphics[width = \columnwidth]{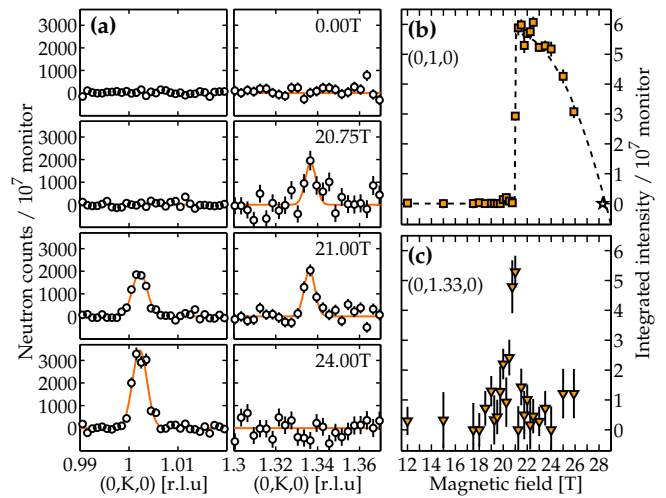}
	\caption{Neutron diffraction results from HFM/EXED. (a) neutron counts as a function of $(0,K,0)$ around $K = 1$ (left panels) and $K = 4/3$ (right panels) at selected field values. The orange lines are Gaussian fits.  (b) and (c) integrated intensity as a function of magnetic fields up to $25.9\,\mathrm{T}$ of $(0,1,0)$ and $(0,4/3,0)$ respectively. The star symbol in (b) shows the expected zero intensity of $(0,1,0)$ at saturation field\cite{kharchenko2010}. The dashed line is a guide to the eye.}
	\label{fig:HFM}
\end{figure}

To access fields approaching the saturation field, $\mu_0 H_S = 28.3\,\mathrm{T}$, a neutron diffraction was performed at the HFM/EXED instrument. The maximum field was $25.9\,\mathrm{T}$ and thus enabled direct probing of the remaining magnetic phases at high fields. The required crystal orientation and the opening angle of the magnet limited the number of accessible Bragg peaks to $(\bar{3},0,\bar{1})$, $(\bar{2},0,\bar{1})$, $(\bar{1},0,\bar{1})$, $(1,0,\bar{1})$ , $(\bar{1},0,0)$, $(0,0,\bar{1})$ and $(0,K,0)$ for $K \lesssim 10$. All peaks except $(0,K,0)$ were observed in the forward scattering detectors and unfortunately, due to low flux at the required wavelengths, the neutron statistics of these peaks were only sufficient for alignment and confirmation of the zero-field structure. However, magnetic intensity above $20.5\,\mathrm{T}$ was observed in the backscattering detectors at the $(0,K,0)$ position. Intensity was found at $K = 4/3$ for $20.5-21.0\,\mathrm{T}$ and at $K = 1$ above $21.0\,\mathrm{T}$ with the two peaks coexisting at $21.0\,\mathrm{T}$. Neutron counts as a function of $K$ along $(0,K,0)$ were obtained by integrating over a slice in reciprocal space of dimensions (given in r.l.u.) $\Delta H = 0.3$ and $\Delta L = 0.2$ and with bin sizes $\Delta K = 1 \times 10^{-3}$ and $\Delta K = 3 \times 10^{-3}$ for $K = 1$ and $K = 4/3$ respectively. Background subtracted line profiles at selected field strengths are shown in Fig. \ref{fig:HFM}(a) and integrated intensities of $(0,1,0)$ and $(0,4/3,0)$ found from Gaussian fits are shown in Fig. \ref{fig:HFM}(b) and (c) respectively.

The ordering vector in the interval $20.5-21.0\,\mathrm{T}$ is thus ${\bf Q} = (0,1/3,0)$. This is the same as for $11.9-20.5\,\mathrm{T}$ but the $(0,4/3,0)$ Bragg peak is not present in the cycloid phase. Although the period of the magnetic structure stays the same, the spin orientation must then change. In Fig. \ref{fig:HFM}(c) the transition appears abrupt but seems continuous in the magnetization data of Ref. \onlinecite{kharchenko2010}. One possibility consistent with these observations is a gradual transition from the cycloid to a conical structure with the cone base perpendicular to the propagation vector. In such a structure the spins rotate in the $(a,c)$-plane and have a ferromagnetic component along the $b$-axis. However, since only a single magnetic Bragg peak was observed a rigorous structure determination is impossible. 

Above $21.0\,\mathrm{T}$ the neutron intensity at $(0,4/3,0)$ vanishes and a new peak appears at $(0,1,0)$. This peak reflects a commensurate spin structure with symmetry $(\uparrow \uparrow \downarrow \downarrow)$, the same as in the zero-field phase where the spins are predominantly along $b$. Since neutron scattering is only sensitive to spin components perpendicular to the scattering vector this Bragg is not observed in the zero-field phase. Conversely, the finite peak intensity above $21.0\,\mathrm{T}$ implies antiferromagnetic spin components along either $a$ or $c$ instead of $b$. Both susceptibility measurements and the magnetic structure refinement in the cycloid phase suggest that the $c$-axis is easier than $a$. Therefore, we infer that above $21.0\,\mathrm{T}$ the major antiferromagnetic spin component is along $c$. In addition, there is a ferromagnetic component with $2/3\,M_S$ at $21.0\,\mathrm{T}$ which increases approximately linearly until saturation is achieved at $28.3\,\mathrm{T}$\cite{kharchenko2010}. The magnetic structure above $21.0\,\mathrm{T}$ may therefore be described as a magnetized spin-flop structure. The spins rotate towards the $b$-axis with increasing field and the intensity of $(0,1,0)$ decreases with field accordingly. In fact, the field dependence of $(0,1,0)$ is consistent with its complete disappearance at saturation, see Fig. \ref{fig:HFM}(b). 

The Bragg peaks at $(0,4/3,0)$ and $(0,1,0)$ coexist in a short field interval, see Fig. \ref{fig:HFM}(a), suggesting that the phase transition from the ${\bf Q} = (0,1/3,0)$ to the commensurate phase is of first order. This is also substantiated by hysteresis observed in previous pulsed-field magnetization measurements\cite{kharchenko2010}.

\section{Discussion}

\subsection{The cycloid structure and a possible magnetoelectric effect}

At first glance the cycloid structure, Fig. \ref{fig:structure}(b)-(c), seems counterintuitive when regarding the axial single-ion anisotropy and antiferromagnetic nearest neighbor interactions. Neither exchange nor the single-ion anistropy energies are minimized. However, the deviations of moments from the $b$-axis remain relatively small such that spins are either nearly antiparallel or parallel. It is also noteworthy that the spins are in the $(b,c)$-plane as opposed to the $(a,b)$-plane, signifying that the energy cost for spins along $c$ is smaller than along $a$ as expected from both susceptibility measurements\cite{creer1970} and density functional theory\cite{yamauchi2010}.

LiCoPO$_4$ has a strong magnetoelectric effect in the commensurate low-field phase\cite{mercier}. Here an electric polarization, $P_a$, is induced along $a$ for magnetic fields applied along $b$ and vice versa. The magnetoeletric properties of the phase with the 1/3 magnetization plateau have also been studied with the conclusion that this phase does not display the same magnetoelectric effect\cite{wiegelmann,khrustalyov2016}. However, from symmetry analysis the cycloid structure does actually support a magnetoelectric effect\cite{kimura2007} but via a different mechanism: the inverse Dzyaloshinskii-Moriya effect. The direction of the allowed electric polarization is along ${\bf k} \times ({\bf S}_i \times {\bf S}_j)$, and in the case of the cycloid in LiCoPO$_4$, ${\bf k} || {\bf b}$ and $({\bf S}_i \times {\bf S}_j) || {\bf a}$. Hence the polarization would be along the $c$-axis for magnetic fields applied along $b$. To our best knowledge only $P_a$ was measured in the previous studies and the allowed component, $P_c$, has not yet been probed. Therefore, the possibility of a magnetoelectric effect in the ${\bf Q} = (0,1/3,0)$ cycloid structure is not definitely rejected and should be further investigated.

\subsection{Hysteresis and stacking faults}

\begin{figure*}
	\includegraphics[width = \textwidth]{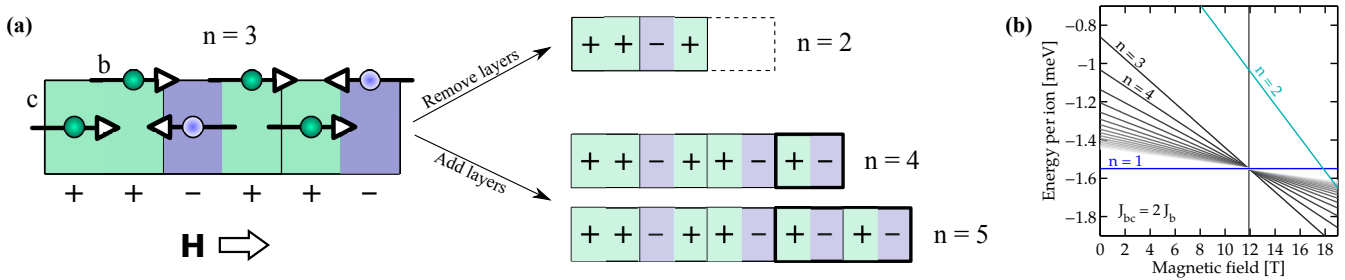}
	\caption{Stacking faults. (a) Possible stacking fault structures with period $n \in \mathbb{N}$ for $n > 1$ and with magnetization $1/n \, M_S$. Spin direction is denoted by the ion color: green (along $b$) and blue (along $\bar{b}$). Only one layer of ions in the $(b,c)$ plane is shown here. (b) Energy per magnetic ion as a function of applied field calculated from the stacking fault structures. $J_{bc} = 2 J_b$ is assumed. The zero-field energy, $E_1$, is shown with the solid blue line and $E_n \rightarrow E_1$ for $n \rightarrow \infty$. The energies for $n > 2$ cross at this level exactly at the transition field $H_C = 11.9\mathrm{T}$ as shown with the vertical line.}
	\label{fig:stacking}
\end{figure*}

The features observed in the magnetization and the occurence of the $(3,1/3,1)$, $(3,1/4,1)$ and $(3,1/2,1)$ incommensurate peaks in the interval $11.4-11.9\,\mathrm{T}$ are consistent with the behavior of the $(3,0,1)$ intensity as a function of applied field seen in the RITA-II experiment. Similarly, field ramp direction dependent differences in the curve shapes of the electric polarization were presented in Ref. \onlinecite{khrustalyov2016}. Hence, extra features in the transition regime are established in several measurable quantities. Upon increasing the temperature, the effects weaken: at $T \gtrsim 6\,\mathrm{K}$ the difference in curve shape in the magnetization is mostly absent and so is the multiple ordering vectors for increasing field as well as the Lorentzian broadening for decreasing field, compare top and bottom panels in both Figs. \ref{fig:mag_rita} and \ref{fig:colorplots}. In the following, a relatively simple model based on a mean-field approach is proposed in an attempt to understand these observations.

The magnetic structure above $11.9\,\mathrm{T}$, Fig. \ref{fig:structure}(b) and (c), provides a starting point for our model. The period of this structure is $n = 3$ (i.e. the size of the magnetic unit cell triples) and spins with the same $y$ coordinate form a layer in the $(a,c)$-plane. The spins of each layer are then rotated with respect to those in the next layer upon advancing along $b$. In the present model, we crudely assume that all moments have maximum length, $M_S = 3.6\mu_B$, and that they are purely oriented along the easy axis. Hence the canting of $\sim 20^{\circ}$ away from the $b$-axis for $2/3$ of the spins is completely ignored here. This structure consists of two kinds of layers or building blocks: (i) layers with spins parallel to $b$ and (ii) layers with spins antiparallel to $b$. Each crystallographic unit cell consists of two such layers. These blocks are denoted ''$+$'' and ''$-$'' respectively and the $n = 3$ structure can then be described by the stacking sequence $[++-++-]$.

Additional magnetic structures are now constructed from the same building blocks such that they have period $n \in \mathbb{N}$ for $n > 1$ and magnetization $1/n \, M_S$. This is done by adding or removing layers of ''$+$'' and ''$-$'' in pairs along $b$. Thus the $n = 4$ structure becomes $[++-++-+-]$, see Fig. \ref{fig:stacking}(a). It can be described by introducing \textit{stacking faults} to the $n = 3$ structure in analogy with stacking faults in hexagonal closed packed structures with layer stacking in e.g. either $ABABAB$ or $ABCABC$ type sequences. 

Note that the constructed structures are not associated with a single $(0,1/n,0)$ ordering vector but require higher harnomics for a full description. However, the associated Bragg peaks are too weak to be detected in our experiment. Furthermore, sufficiently large domains of a structure of period $n$ must exist in the sample in order to observe an $(0,1/n,0)$ ordering vector. At this point it should also be emphasized that the proposed model is not the outcome of a full statistical treatment but rather the proposed stacking fault structures are deliberally chosen to be consistent with experimental observations. It is therefore fully possible that other choices yield similar results. Nevertheless, as we shall see below, this rather crude model provides an explanation of the observed coexistence of several propagation vectors in the transition region.

To describe the energy of the system the following Hamiltonian is employed:
\[
	\mathcal{H} = -\sum_{i,j} J_{ij} {\bf S}_i \cdot {\bf S}_j - \mu H \sum_i S_i,
\]
where only $J_{bc}$ and $J_b$ are taken into account as the remaining exchange constants are generally small in the lithium orthophosphate family\cite{jensen2009,toftpetersen2015,tian2008}. Since the easy axis is along $b$ and the assumed spin structures have no components along $a$ or $c$, no single-ion anisotropy terms are taken into account. $H$ is the strength of the applied field along $b$ and $\mu = g \mu_B S$ with the gyroscopic ratio $g \approx 2$, the Bohr magneton $\mu_B$ and $S = 3/2$. The energy per Co$^{2+}$ ion of the assumed stacking fault structures with period $n$ is then:
\begin{align*}
	& E_n = \frac{1}{n} \Big( \left[ 2(n-2)J_{bc} + (4-n)J_b \right] S^2 - \mu H \Big),\\
	&	\hspace{5.5cm} n > 2, \ n \in \mathbb{N}.
\end{align*}
The zero-field structure, i.e. $n = 1$, see Fig. \ref{fig:phasediagram}, has the energy per ion $E_1 = (2J_{bc} - J_b) S^2$. By solving $E_1 = E_n$ one can determine the transition field from the zero-field structure to any stacking fault structure accordingly:
\begin{align*}
	& H_C = \frac{4 S^2}{\mu} \Big( -J_{bc} + J_b \Big), \quad n > 2, \ n \in \mathbb{N}.
\end{align*}
Peculiarly, the transition field is independent of the period $n$ and hence all configurations of this particular kind are degenerate exactly at the phase transition. The energy difference between any two states $m$ and $n$ is readily calculated:
\begin{align*}
	& E_m - E_n = \left( \frac{m - n}{nm} \right) \mu \ (H - H_C), \quad n > 2, \ n \in \mathbb{N}.
\end{align*}
Hence, the energy difference does not depend directly on exchange interactions but merely on $m$ and $n$ as well as the field deviation from the transition value.

A short note on the $n = 2$ state is in place since the above calculations are only valid for $n > 2$. For $n = 2$ the stacking sequence results in a different expression for the energy, $E_2 = - \frac{1}{2} \mu H$, and a larger transition field follows. It is therefore unlikely that this structure is realized. Alternatively, the $n = 2$ Bragg peak could be due to nuclear distortion linked to the $n = 4$ magnetic structure or simply a completely different magnetic structure with period $n = 2$. An X-ray or polarized neutron experiment is needed in order to clarify this point.

Assuming $J_{bc} \approx 2J_b$ and using the measured transition field of $11.9\,\mathrm{T}$ the nearest neighbor coupling strength is estimated to $J_{bc} \approx -0.46\,\mathrm{meV}$. With these assumptions, energies for different $n$ configurations are shown as a function of applied field in Fig. \ref{fig:stacking}(b). The estimate of the relative strengths of $J_{bc}$ and $J_b$ is based on the other members of the lithium orthophosphate family\cite{jensen2009,toftpetersen2015,yiu2017,li2009}. The resulting value of the nearest neighbor interaction is remarkably close to those found in LiFePO$_4$ ($J_{bc}=-0.46(2)\,\mathrm{meV}$\cite{yiu2017}) and LiMnPO$_4$ ($J_{bc}=-0.48(5)\,\mathrm{meV}$\cite{li2009}) and reasonably close to that measured for LiCoPO$_4$ ($J_{bc}=-0.7(2)\,\mathrm{meV}$\cite{tian2008,Note1}, note the large uncertainty). Additionally, in Ref. \onlinecite{kharchenko2010} the nearest neighbor interaction of LiCoPO$_4$ was estimated to $J_{bc} = -0.23\,\mathrm{meV}$. However, this result is based on an incorrect magnetic structure explaining the discrepancy from our result. It is worth emphasizing here that our result is obtained merely from a few simple but reasonable assumptions together with the measured transition field value. 

It is clear from Fig. \ref{fig:stacking}(b) that the energy difference between different $m$ and $n$ states is small close to the transition field. Hence, at low temperatures the thermal relaxation time may be sufficiently long such that regions of the sample are trapped in states with $n \neq 3$ in agreement with the observation of $n = 4$ order in the transition interval, $11.4-11.9\,\mathrm{T}$. At $11.9\,\mathrm{T}$ the $n = 3$ structure stabilizes and the other structures withdraw. At higher temperatures the system is assisted by thermal fluctuations and rapidly finds its stable configuration. Thus, based on the disappearance of hysteresis at higher temperatures ($T\gtrsim 6 \,\mathrm{K}$) it is suggested that more states, e.g. $n = 5, 6$, may be populated at very low temperatures (mK regime) and that the hysteresis region is significantly expanded. Further experiments are needed in order to falsify or substantiate this hypothesis.

For decreasing field the transition to the low-field antiferromagnetic ground state occurs abruptly even at low temperatures. A broadening of the commensurate Bragg peak is observed in a short field interval at the transition. Although very speculative, it may be suggested to originate from long wavelength stacking fault structures like those introduced above, i.e. for $n>>1$. When a large number of $[+-]$ layer pairs are added the magnetization approaches zero and the structure resembles the zero-field structure.

The reason for the transition to occur more readily for decreasing field as compared to increasing field remains unexplained but an analogy may be found in the water solid-liquid transition. Upon heating water ice it slowly melts when the temperature is above $0^{\circ}$C. However, because of the need for nucleation sites, upon cooling, liquid water can reach temperatures below the freezing temperature (supercooling) before suddenly entering the ice phase. In this analogy heating corresponds to increasing field.

\subsection{Commensurability and magnetoelectric effect}

Although only a single magnetic peak was observed above $21.0\,\mathrm{T}$ it is possible to argue that the magnetic structure here is a commensurate, magnetized spin-flop structure with the same main antiferromagnetic symmetry component as the zero-field structure. Remarkably, this phase is magnetoelectric as was recently measured by Kharchenko et al.\cite{khrustalyov2016}. Here an electric polarization, $P_a$, is induced along the $a$-axis for a magnetic field applied along $b$. Thus the active magnetoelectric tensor element, $\alpha_{ab}$, is the same as in the low-field phase but $\sim 5$ times weaker. Such reentrant magnetoelectric effect has previously been observed in the sister compound LiNiPO$_4$\cite{toftpetersen2017,khrustalyov2016}. Here an extension of the microscopic model explaining the low-field effect succeeds in accounting for the high-field effect too. In LiCoPO$_4$ there is of yet no such microscopic model but the two compounds have one characteristic in common: the magnetoelectric effect is linked to commensurate magnetic structures. This is interesting since other magnetoelectric materials such as Cr$_2$BeO$_4$\cite{newnham1978} and {\it R}Mn$_2$O$_5$\cite{nakamura1997} ({\it R} = rare earth) generally display incommensurate magnetic structures\cite{kimura2007}. However, when recalling the above discussion on the possibility of a magnetoelectric effect in the cycloid structure, it appears that LiCoPO$_4$ may support a magnetoelectric effect for both commensurate and incommensurate structures. If this is the case, the effects are most likely caused by two different mechanisms.

\section{Conclusions}

We studied the phase diagram of LiCoPO$_4$ for fields up to $25.9\,\mathrm{T}$ applied along $b$ using magnetization measurements and neutron diffraction. The magnetic structure for $11.9-20.5\,\mathrm{T}$ was determined. The ordering vector is ${\bf Q} = (0,1/3,0)$, demonstrating a tripling of the magnetic unit cell in the $b$-direction. The spin configuration is an elliptic cycloid with spins in the $(b,c)$-plane in superposition with a ferromagnetic component. The ratio of the major and minor axes is $3.2(5)$ with the major axis along $b$. The resulting structure has the spin direction alternating with $2/3$ of the spins almost parallel to and $1/3$ antiparallel to the field, consistent with the observed $1/3$ magnetization plateau\cite{kharchenko2010}. This structure maintains the axial single-ion anisotropy character of LiCoPO$_4$. Furthermore, the refined structure allows for the magnetoelectric effect with an electric polarization induced along $c$ for magnetic fields applied along $b$. The existence of this effect is still to be rejected or confirmed by further measurements.

The transition from the low-field to the cycloid phase exhibits hysteresis and the way the transition proceeds depends heavily on the field ramp direction. For increasing field, we have evidence for three coexisting propagation vectors, ${\bf Q} = (0,1/4,0)$, ${\bf Q} = (0,1/3,0)$ and ${\bf Q} = (0,1/2,0)$, in the field interval $11.4\,\mathrm{T}-11.9\,\mathrm{T}$. The occurence of the additional ordering vectors may be rationalized by introducing \textit{stacking faults} in the cycloid structure leading to states sufficiently close in energy to be populated until a single phase stabilizes at $11.9\,\mathrm{T}$. For decreasing field the transition is more abrupt and the commensurate peak has a Lorenzian lineshape at the transition.

We also determined the ordering vectors in the phases at $20.5-21.0\,\mathrm{T}$ and above $21.0\,\mathrm{T}$. The former has propagation vector ${\bf Q} = (0,1/3,0)$ but a different spin orientation compared to the cycloid phase. The latter is commensurate with a ferromagnetic component along $b$ as well as an antiferromagnetic component along $c$.

\section*{Acknowledgements}

We greatly acknowledge Niels Hessel Andersen for many fruitful discussions and Oleg Rivin, Xinzhi Liu, Robert Wahle and Sebastian Gerischer for their support at the HFM/EXED facility at the Helmholtz-Zentrum Berlin. Work was supported by the Danish Agency for Science, Technology and Innovation under DANSCATT. Neutron experiments were performed at the Swiss spallation neutron source SINQ at the Paul Scherrer Institute, Villigen, Switzerland, at the research reactor at the Institut Laue-Langevin, Grenoble, France and at the BER II research reactor at the Helmholtz-Zentrum Berlin, Germany. Research at Ames Laboratory is supported by the U.S. Department of Energy, Office of Basic Energy Sciences, Division of Materials Sciences and Engineering under Contract No. DE-AC02-07CH11358.



\end{document}